\documentclass[onecolumn,amsmath,amssymb,12pt,superscriptaddress,nofootinbib,floatfix]{revtex4}
\pdfoutput=1

\usepackage{hyperref}
\usepackage{epsfig}
\usepackage{amsmath,amssymb}
\usepackage{verbatim}

\newcommand{\be}{\begin{equation}}
\newcommand{\ee}{\end{equation}}
\def\bea#1\eea{\begin{align}#1\end{align}}

\renewcommand{\d}{\textrm{d}}

\newcommand{\SL}{\mathop{\rm SL}}
\newcommand{\SO}{\mathop{\rm SO}}

\begin{document}

\title{Axion Wormholes in AdS Compactifications}
\author{Thomas Hertog}
\email{thomas.hertog@kuleuven.be}
\affiliation{Institute for Theoretical Physics, KU Leuven, 3001 Leuven, Belgium}
\author{Mario Trigiante}
\email{mario.trigiante@polito.it}
\affiliation{Department of Applied Science and Technology, Politecnico di Torino, \\ C.so Duca degli Abruzzi, 24, I-10129 Torino, Italy}
\author{Thomas Van Riet}
\email{thomas.vanriet@kuleuven.be}
\affiliation{Institute for Theoretical Physics, KU Leuven, 3001 Leuven, Belgium}

\begin{abstract}
\vspace{1cm}

We find regular axionic Euclidean wormhole solutions in Type IIB string theory compactified on $\text{AdS}_5\times \text{S}^5/\mathbb{Z}_k$. AdS/CFT enables a precise derivation of the axion content of the Euclidean theory, placing the string theory embedding of the wormholes on firm footing. This further sharpens the paradox posed by these solutions.

\end{abstract}

\pacs{XXX}

\maketitle

\section{Introduction}

Euclidean wormholes \cite{Giddings:1987cg,Lavrelashvili:1987jg,Hawking87} are extrema of the action in Euclidean quantum gravity that connect two distant regions, or even two disconnected asymptotic regions. Despite much work over many years it remains unclear whether wormholes can provide valid saddle point contributions to the Euclidean path integral and therefore have physical implications (see e.g. \cite{Coleman:1988cy, Bergshoeff:2005zf, ArkaniHamed:2007js, Bergman:2007ss,Giddings88}). 

The Weak Gravity Conjecture (WGC) \cite{ArkaniHamed:2006dz} adds a new dimension to this question. This is because its generalization to instantons implies the existence of super-extremal instantons which, when sourced by axions, correspond to Euclidean axion wormholes. It has been argued that such instanton contributions can destroy the flatness of the potential in models of large field inflation based on axions \cite{Montero:2015ofa}, although there is no consensus on this \cite{Hebecker:2016dsw}. 

It is therefore important to elucidate the physical meaning - if any - of wormholes. To this end it is clearly of interest to find wormhole solutions in string theory and in particular in AdS compactifications, where the AdS/CFT dual partition function provides an alternative description of the gravitational path integral. Axionic wormholes \cite{Giddings:1987cg} provide natural candidates for wormhole solutions in string theory. However axions are always accompanied by dilatons in string theory compactifications, and the existence of regular wormhole solutions depends delicately on the number of scalars and their couplings \cite{ArkaniHamed:2007js}. In a single axion-dilaton system coupled to gravity, for instance, the dilaton coupling must be sufficiently small in order for wormholes to exist. 

In \cite{Bergshoeff:2004pg} Calabi-Yau compactifications were found which allow for regular axionic wormhole solutions in flat space. The situation is more subtle however in  compactifications to AdS. Type IIB on $AdS_5 \times S^5$ does not admit axionic wormholes \cite{Bergshoeff:2005zf}. On the other hand, in \cite{ArkaniHamed:2007js} it was argued there are approximate wormhole solutions in Type IIB compactified on $AdS_3\times S^3\times \mathbb{T}^4$. However no clean derivation was given to determine the exact axion - dilaton content of this compactification\footnote{In this paper the word axion refers to scalar fields with shift symmetries whose kinetic term has flipped sign in the Euclidean theory. All other scalars will be referred to as dilatons.}. The validity of those solutions therefore remains somewhat uncertain. Specifically, their smoothness depends on the specific Wick rotation that was used in \cite{ArkaniHamed:2007js}, but it remains unclear whether this particular Wick rotation is the one selected by AdS/CFT.

The goal of this paper is to construct exact, regular axionic wormhole solutions in an AdS compactification where the Wick rotation to the Euclidean theory can be made rigorous using AdS/CFT. The wormholes we find are solutions to Euclidean IIB string theory on $\text{AdS}_5\times \text{S}^5/\mathbb{Z}_k$, whose field theory duals are certain $\mathcal{N}=2$ quiver theories \cite{Kachru:1998ys}. The dual operators that are turned on are exactly marginal operators, which enables us to identify the Wick rotation selected by AdS/CFT and therefore rigorously determine the nature of the scalar fields in the theory.
Our results further sharpen the paradox with AdS/CFT and the apparent uniqueness of quantum gravity. One is left wondering what is pathological about axionic wormholes.

\section{Axion Wormholes}
To discuss wormhole solutions in a string theory setting we consider a Euclidean theory in $D$ dimensions consisting of gravity coupled to massless scalars $\phi^I$ and a negative cosmological constant $\Lambda \equiv -(D-1)(D-2)/2l^2$,
\begin{equation}
S=\int\sqrt{g}\Bigl(R -
\tfrac{1}{2}G_{IJ}\partial\phi^I\partial\phi^J - \Lambda\Bigr)\,,
\end{equation}
where $G_{IJ}$ is a general sigma model metric. Euclidean Lagrangians of this kind describe, for instance, certain consistent truncations of AdS compactifications of string theory. In this setup the scalars live in the moduli space of the AdS vacuum and the metric $G_{IJ}$ corresponds to the Zamolodchikov metric on the conformal manifold of the dual CFT. In the Euclidean theory this metric does not need to have a definite signature (see the discussion around equation (\ref{2point}). In particular it typically contains dilatonic scalars $\varphi$ which enter with kinetic terms with the usual sign, as well as {\it axionic} scalars $\chi$ which have the `wrong' sign kinetic terms. The presence of axions is crucial indeed for the theory to admit regular wormhole solutions.

We are interested in wormhole solutions described by the following $\SO(D)$ invariant metric,
\begin{equation} \label{wormholegeometry}
\d s^2 = f(r)^2\d r^2 + a(r)^2\d \Omega^2\,.
\end{equation}
The equations of motion for the scalars in geometries of this form reduce to geodesic equations in the metric $G_{IJ}$. Therefore, in an affine parametrisation, we have
\begin{equation}
G_{IJ}\dot{\phi}^I\dot{\phi}^J = c\,.
\end{equation}
where $\dot \phi = a^{D-1}d\phi/dr$.
Inserting this in the constraint equation for $a$ yields
\begin{equation}\label{firstorder}
f^{-2}(r)\left(\frac{\d}{\d r}a\right)^2 = \frac{c}{2(D-1)(D-2)}a^{4-2D} +
\frac{a^2}{l^2} + 1\,.
\end{equation}
The nature of the solutions depends on the sign of $c$. When $c=0$ we have $a(r)=l\sinh(r/l)$ and the solution is AdS, for $c>0$ the solutions are singular ``cored instantons'' but when the geodesic motion on moduli space is timelike, corresponding to $c<0$, regular wormhole solutions are possible\footnote{In terms of BPS bounds $c>0$ is sub-extremal, $c=0$ is extremal and $c<0$ is super-extremal. The role of mass is played by on-shell action and charge is defined with respect to the scalar fields. For axion fields, the latter is the charge under a global shift symmetry. In general it is a charge under the isometries of the scalar manifold.}.


Since only the axionic fields give negative contributions to
$G_{IJ}\dot{\phi}^I\dot{\phi}^J$, solutions with $c<0$, and thus
wormholes, only exist in theories that have axions. Axions arise naturally in Euclidean theories with $D-2$-form fields $B$. Consider for instance the following theory
\begin{equation}
S=\int\sqrt{g}\Bigl(R -\tfrac{1}{2}(\partial\varphi)^2 -
\frac{1}{2 (D-1)!}e^{-b\varphi} |\d B|^2 - \Lambda\Bigr)\,,
\end{equation}
with $b$ a real number, called the dilaton coupling. After Hodge dualising the form field to a scalar $\chi$ the effective action becomes
\begin{equation}\label{action}
S=\int\sqrt{g}\Bigl(R -\tfrac{1}{2}(\partial\varphi)^2 +
\tfrac{1}{2}e^{b\varphi}(\partial \chi)^2 - \Lambda\Bigr)\,.
\end{equation}
and hence $\chi$ is an axion. Thus if it can be argued that the $B$-form is fundamental in the Lorentzian theory, then this yields a Euclidean theory with an axion upon Wickrotation. In the absence of this, however, it has proven difficult to determine unambiguously whether scalars enter as dilatons or axions in Euclidean supergravity theories arising e.g. in string theory compactifications\footnote{Note that supersymmetry does not resolve this, since it does not uniquely determine the signs of the kinetic terms in the scalar Lagrangian in the Euclidean theory.}. One of the advantages of the setup we consider below, is that we can use AdS/CFT to specify unambiguously the dilaton - axion content of the theory.

A more precise condition on the moduli space metric $G_{IJ}$ in order for the theory to admit regular wormhole solutions follows from the constraint \eqref{firstorder}. This implies that the geodesic distance $d$ between the values of the scalars at opposite ends of a regular wormhole solution is given by \cite{ArkaniHamed:2007js}
\begin{equation}\label{length}
d= \int_{r=-\infty}^{r=+\infty}\sqrt{|G_{ij}\dot{\phi}^i\dot{\phi}^j|\d r }= 2\int^{a_0}_{-\infty}|c|a^{2(1-D)}\frac{\d a}{\dot a} 
\approx \pi\sqrt{2\frac{D-1}{D-2}} - {\cal O}\left(a_0/l\right)
\end{equation}
where $a_0$ denotes the minimum size of the wormhole neck and the first term is the result for $\Lambda=0$. The leading correction in $a_0/l$ enters with a negative sign but this is  small when the wormholes are small relative to the AdS scale.



Thus, in order for there to be a regular wormhole solution there must be a compact timelike geodesic at least as long as \eqref{length} in the scalar moduli space. For a single axion-dilaton system the maximal length $d_{max}$ of geodesics on the moduli space manifold $\SL(2, \mathbb{R}/\SO(1,1)$ is
\be
d_{\text{max}}= \frac{2\pi}{|b|}\,.
\ee
which leads to the following condition for the existence of smooth axion-dilaton wormholes in flat space,
\be\label{cond}
|b|<\sqrt{\frac{2(D-2)}{D-1}}\,.
\ee
Eq. \eqref{length} shows this requirement is slightly weaker in AdS.

It may seem difficult to obtain general conclusions about the length of geodesic curves on more general moduli spaces. However, the manifolds of the moduli spaces in the extended supergravity theories of interest here are often symmetric cosets $G/H$. In these theories it was shown \cite{Bergshoeff:2008be} that many properties of geodesics curves can be deduced from the study of  seed geodesics. Seed geodesics are the minimal set of geodesics that generate all geodesics by acting with the isometry group. In particular it was found that the longest timelike seed geodesics live in a direct product of $p$ axion-dilaton pairs \cite{Bergshoeff:2008be}:
\be \label{theorem}
\text{seed geodesic}\quad \in \quad \Bigr[\frac{\SL(2, \mathbb{R})}{\SO(1,1)}\Bigl]^p
\ee
where $p$ depends on the details of the coset and on the Wickrotation. The maximal length geodesic on the symmetric coset is then given by
\be
d_{\text{max}}= \frac{2\pi}{|b_\text{eff}|}\,\quad\text{with}\,\quad
\frac{1}{b_{\text{eff}}^2} \equiv \sum_{i=1}^p \frac{1}{b_i^2}\,.
\ee
where $b_i$, with $i=1\ldots p$, are the coupling constants of the individual axion - dilaton pairs in the theory. Therefore, even though some of the individual coupling constants may not satisfy the condition \eqref{cond}, as long as there are sufficiently many axion - dilaton pairs the theory will admit wormhole solutions \cite{ArkaniHamed:2007js}. 

\section{A Remark on Wick rotations}

As mentioned earlier, it has proven difficult to specify unambiguously the axion - dilaton content of Euclidean supergravity theories. This has hindered attempts to provide clean embeddings of wormhole solutions in AdS compactifications.

We now discuss two rigorous methods to construct Euclidean theories with axionic scalars. The first approach uses dimensional reduction over time. We explain why this does not yield theories admitting regular wormholes. The second approach makes use of AdS/CFT, and we use this to find new wormhole solutions in string theory in Section \ref{sol}.

\vspace{.1in}\noindent
{\it Timelike reduction}\\
Consider a general truncation of a Lorentzian (super-)gravity theory in four dimensions containing gravity, coupled to scalars and vectors. Compactifying one of the space dimensions on a circle yields a theory in three dimensions in which all vectors can be dualised to scalars. Consider the case in which the scalars then parameterize a symmetric coset manifold $G/H$, where $H$ is the maximal compact subgroup of $G$. This is the case for all supergravity theories with at least 16 supercharges and many examples with less supercharges exist as well. 

Now instead compactify the original Lorentzian theory on the time direction. This yields a Euclidean theory in three dimensions, again with no vectors. The scalars now parameterize a pseudo-Riemannian coset $G/H^*$ where $H^*$ is a specific non-compact subgroup of $G$  \cite{Breitenlohner:1987dg, Bergshoeff:2008be}. Thus we obtain a specific Wick rotation of $G/H$:
\be
\frac{G}{H}\quad\underset{\text{Wickrotation}}{\implies} \quad \frac{G}{H^*}\,.
\ee
This Wickrotation can be understood as follows: All vectors that are being dualised to scalars after time-like reduction, as well as all scalars that come from reducing the vector (the temporal components of vectors in 4D), enter as axions. All other scalars are dilatons. Physically the axionic scalars correspond to the magnetic and electric fields of four dimensional black holes. 

A well-known example of this construction is the reduction of maximal supergravity to three dimensions where the Wick rotation corresponds to
\be
\frac{E_{8(+8)}}{\SO(16)}\quad\underset{\text{Wick rotation}}{\implies} \quad \frac{E_{8(+8)}}{\SO^*(16)}\,,
\ee
Interestingly, there are no models $G/H^*$ constructed in \cite{Bergshoeff:2008be} for which the condition \eqref{cond} for wormholes to exist, is satisfied. This can be understood as follows: Imagine that regular wormholes existed in these theories. This would mean, using the oxidation formulae in \cite{Breitenlohner:1987dg} (see also \cite{Bergshoeff:2008be, Bossard:2009at}), there are smooth, four-dimensional black holes in the original theory that violate the generalised Reissner-Nordstrom bound. However this was proven to be impossible \cite{Breitenlohner:1987dg}. 

By contrast, if one were to Wick rotate liberally -- that is, with mathematical consistency as the only guiding principle but without a corresponding `physical' reduction --  it is easy to construct theories with regular wormholes. An example was given in \cite{Bergshoeff:2008be} using the following Wick rotation,
\be
\frac{E_{8(+8)}}{\SO(16)}\quad \underset{\text{Wick rotation}}{\implies}\quad \frac{E_{8(+8)}}{\SO(8,8)}\,,
\ee

However it appears plausible that the Wick rotation in string theory compactifications is further constrained and in particular excludes constructions of this kind. We now turn to a different approach, based on AdS/CFT, to construct Euclidean theories. 

\vspace{.1in} \noindent
{\it Using holography}\\

AdS/CFT maps the AdS moduli space to the conformal manifold of the dual field theory. The latter is naturally endowed with the Zamolodchikov metric \cite{Zamolodchikov:1986gt}, defined via the two point functions $\langle \mathcal{O}_i \mathcal{O}_j \rangle $ between marginal operators $\mathcal{O}_i$. There is a natural way to Wick rotate this metric: If the marginal operator acquires an $i$-factor in the Lagrangian in Euclidean signature, then the metric in that direction must change signature so that the 2-point function is defined with an extra $i^2$ maintaining consistence with the dual. 

An example to keep in mind is the operator $\text{Tr}F\wedge F$ in Yang-Mills theory that couples to the $\theta$-angle, where we have
\be \label{2point}
\theta \text{Tr}F\wedge F \quad \underset{\text{Wickrotation}}{\implies}\quad i\theta \text{Tr}F\wedge F \,,
\ee
because  $\text{Tr}F\wedge F$ is a pseudo-scalar.  For $\mathcal{N}=4$ SYM the holographic dual to $\theta$ is the boundary value of the RR axion $\chi$. Consistently, the RR axion is Wick rotated in Euclidean IIB supergravity \cite{Belitsky:2000ws} since the axion can be argued to be a pseudo-scalar. 

Hence AdS/CFT leads to the following procedure for Wick rotation: a shift-symmetric scalar in the AdS moduli-space Wick rotates whenever it couples to a marginal operator in the dual CFT that is a pseudo-scalar. If one were to adopt a different prescription then the holographic computation of one-point functions would no longer be consistent with known field theory results (see for instance \cite{Balasubramanian:1998de} in case of $\mathcal{N}=4$ SYM). We now apply this procedure to find new axionic wormholes in AdS compactifications.

\section{Axion wormholes in Type IIB on $AdS_5\times S^5/\mathbb{Z}_k$} \label{sol}

Starting with the moduli space of $AdS_5\times S^5/\mathbb{Z}_n$ \cite{Corrado:2002wx,Louis:2015dca}:
\begin{equation}
\mathcal{M}=\frac{{\rm SU}(1,n)}{{\rm S[U(1)\times U(n)]}}\,,
\end{equation}
we now apply the procedure above to derive a Wick rotated theory containing regular wormhole solutions.

The special K\"ahler manifold $\mathcal{M}$ is described by $n$ complex scalar fields $z^a=(u,v_i)$, $i=1,\dots ,n-1$. Its geometry is characterized by a prepotential of the form
$$\mathcal{F}(u,v_i)=\frac{i}{4}\,\left(1-u^2-\sum_{i=1}^{n-1} v_i^2\right)\,.$$
The metric features $n$ commuting translational isometries which define a maximal abelian subalgebra of the isometry algebra $\mathfrak{su}(1,n)$. To describe these transformations and the axionic scalars that are shifted under them, it is useful to introduce a solvable Lie algebra parametrization, in which $\mathcal{M}$ is globally described as isometric to a solvable group manifold generated by a solvable Lie algebra $Solv$: $\mathcal{M}\sim \exp(Solv)$. The manifold is then spanned by scalar fields $U,a,\zeta^i,\,\tilde{\zeta}_i$ which parametrize the generators $H_0,\,T_\bullet,\,T_i,\,T^i$ of $Solv$, whose algebraic structure is defined by the following commutation relations:
\begin{align}
&[H_0,T_M]=\frac{1}{2}\,T_M\,\,,\nonumber\\
&[H_0,T_\bullet]=T_\bullet\,\,,\nonumber\\
&[T_M,\,T_N]=\mathbb{C}_{MN}\,T_\bullet\,,
\end{align}
where we have used the following symplectic notation $T_M\equiv(T_i,\,T^i)$ and $\mathbb{C}_{MN}$ is the ${\rm Sp(2n-2)}$-invariant matrix: 
\begin{equation}
\mathbb{C}\equiv \left(\begin{matrix}{\bf 0} & {\bf 1}\cr -{\bf 1} & {\bf 0}\end{matrix}\right)\,.
\end{equation}
We see that $Solv$ contains a characteristic Heisenberg subalgebra of isometries generated by $T_M,\,T_\bullet$. The geometry of the manifold can be described by the following coset representative in $\exp(Solv)$:
\be
\mathbb{L}=\exp(-a T_\bullet)\,\exp(\sqrt{2}\,\mathcal{Z}^M T_M)\,\exp(2\,U\,H_0)\,,
\ee
where $\mathcal{Z}^M\equiv (\zeta^i ,\,\tilde{\zeta}_i )$. The metric on the coset is then defined as
\begin{equation}
\d s^2 = -\tfrac{1}{2}\text{Tr}[\d M \d M^{-1}]\,,
\end{equation}
where $M = \mathbb{L}\mathbb{L}^{\dagger}$. This gives:
\begin{equation}
ds^2=4 dU^2+{e^{-4U}}\mathcal{N}^2+2{e^{-2U}}\sum_{i=1}^{n-1}[(d\zeta^i)^2+(d\tilde{\zeta}_i)^2]\,,
\end{equation}
having defined $\mathcal{N}\equiv da+\mathcal{Z}^M\mathbb{C}_{MN}d\mathcal{Z}^N$. Inspired by \cite{Andrianopoli:2012ee} we found the following relation between the solvable parameters and the complex coordinates $u,\,v_i$,
\begin{align}
u&=\frac{1-\mathcal{E}}{1+\mathcal{E}}\,\,,\,\,\,v_i\equiv \sqrt{2}\frac{\zeta^i -i\,\tilde{\zeta}_i }{1+\mathcal{E}}\,,\nonumber\\
\mathcal{E}&\equiv e^{2U}+\frac{1}{2}\,\sum_{i=1}^{n-1}[(\zeta^i)^2+(\tilde{\zeta}_i)^2]+i\,a\,.
\end{align}

The Heisenberg algebra of isometries acts by means of the following infinitesimal transformations: $\mathcal{Z}^M\rightarrow \mathcal{Z}^M+\xi^M,\,a\rightarrow a+\beta-\xi^M\,\mathbb{C}_{MN}\mathcal{Z}^N$. The isometries $T^i,\,T_\bullet$ are commuting translations and generate a maximal abelian subalgebra of $\mathfrak{su}(1,n)$ as well as a maximal abelian ideal of $Solv$. Its parameters $\tilde{\zeta}_i,\,a$ are Peccei-Quinn scalars multiplying the $F\wedge F$ terms in the Lagrangian. This therefore tells us how to Wick rotate the sigma model metric. 

In order to understand the Wick rotation it is useful to describe the ${\rm SU(1,n)}$-generators in the fundamental representation, in terms of $(n+1)\times (n+1)$ pseudo-unitary matrices preserving the metric $\eta={\rm diag}(-1,+1,\dots,+1)$. The solvable generators have the following explicit expression:
\begin{align}
H_0&=\frac{1}{2}\,(e_{1,n+1}+e_{n+1,1})\,,\nonumber\\
T_i&=\frac{1}{2}\,(e_{i+1,n+1}-e_{n+1,i+1}-e_{1,i+1}-e_{i+1,1})\,,\nonumber\\
T^i&=\frac{i}{2}\,(e_{i+1,1}-e_{1,i+1}-e_{n+1,i+1}-e_{i+1,n+1})\,,\nonumber\\T_\bullet &=\frac{i}{2}\,(e_{1,n+1}-e_{n+1,1}+e_{n+1,n+1}-e_{1,1})\,,\label{solex}
\end{align} 
where $e_{ab}$ denotes the matrix with all zero entries except for a $1$ in the $a^{th}$ row and $b^{th}$ column.
The Wick rotation $\tilde{\zeta}_i \rightarrow i\,\tilde{\zeta}_i $ and $a\rightarrow i\,a$ amounts to rescaling by an $i$ the complex generators $T^i,\,T_\bullet$ making them real. The resulting Wick-rotated solvable Lie algebra now describes the following symmetric manifold:
\begin{equation}
\mathcal{M}_E=\frac{{\rm SL}(n+1)}{{\rm GL}(n)}\,.
\end{equation}
In other words the Wick rotation, at the level of the isometry algebra, amounts to rescaling by an $i$  the imaginary parts of  the $\mathfrak{su}(1,n)$-generators in the fundamental representation, thus rotating the algebra into $\mathfrak{sl}(n+1)$.
The metric on $\mathcal{M}_E$ reads:
\begin{equation}
ds^2=4 dU^2-{e^{-4U}}\mathcal{N}^2+2{e^{-2U}}\sum_{i=1}^{n-1}[(d\zeta^i)^2-(d\tilde{\zeta}_i)^2]\,.
\end{equation}
This manifold is para-K\"ahler and its coset space has $n$ compact and $n$ non-compact generators.  It has two distinct ${\rm SL}(2)/{\rm SO}(1,1)$ (totally geodesic) submanifolds: one obtained by truncating the sigma-model to $a,U$ and the other to $\psi,\,U$ where $\psi$ denotes one of the $\tilde{\zeta}_i$ axions, say $\psi=\tilde{\zeta}_1$. The metric of the former submanifold is characterized by a value $b=2$ whereas the latter has $b=1$ and thus contains the longest geodesics, with  maximum length $d_{{\rm max}}=2\pi/b=2\pi$. Therefore regular wormholes exist in this AdS compactification. Furthermore, the seed geodesics all live in the second ${\rm SL}(2)/{\rm SO}(1,1)$ subcoset that contains the regular wormholes.  For the special case $n=2$ the BPS instantons supported by this manifold were studied in \cite{Theis:2002er} when $\Lambda=0$.

\section{VII. Discussion}
We have shown that the singular wormholes in $AdS_5\times S^5$ can be turned into everywhere regular wormhole solutions by orbifolding the $S^5$. This is because the orbifolding leads to additional axions in the twisted sector, which makes possible smooth wormholes. With $AdS_5\times S^5$ boundary conditions, the existence of regular wormhole solutions would not be consistent with AdS/CFT since the 1-point function, $\langle |F -\star F|^2 \rangle$, in $\mathcal{N}=4$ SYM computed using holography would be negative \cite{Bergshoeff:2005zf}. It would be very interesting to understand whether the wormhole solutions we have found here lead to an inconsistency of this kind, at the level of the one-point functions in the dual $\mathcal{N}=2$ CFT's. This appears plausible because the fact that extremal (BPS) instantons have lowest action in the field theory most likely implies that the one-point functions computed in the wormhole background violate the Cauchy-Schwarz inequality. 
If so, a possible resolution of this paradox might be that the wormholes have negative eigenmodes \cite{Maldacena:2004rf,Rubakov:1996cn}, which would indicate they don't contribute as saddle points  in the Euclidean path integral defined by the dual. We will report on an analysis of this elsewhere \cite{Hertog17}.

\section{Acknowledgements}

We thank N.~Bobev, H.~Triendl and B.~Vercnocke  for useful discussions. The work of TVR is supported by the FWO odysseus grant G.0.E52.14N. The work of TH is supported in part by the National Science Foundation of Belgium (FWO) grant G092617N, by the C16/16/005 grant of the KULeuven and by the European Research Council grant no. ERC-2013-CoG 616732 HoloQosmos. We also acknowledge supported of the COST Action MP1210 `The String Theory Universe'. We thank the GGI institute in Firenze and the organizers of the workshop `Supergravity:  what next?' for their hospitality.

\appendix*
\setcounter{equation}{0}


\begin{thebibliography}{100}

 \bibitem{Giddings:1987cg}
  S.~B.~Giddings and A.~Strominger,
  Nucl.\ Phys.\ B {\bf 306} (1988) 890.
  doi:10.1016/0550-3213(88)90446-4

\bibitem{Lavrelashvili:1987jg} 
  G.~V.~Lavrelashvili, V.~A.~Rubakov and P.~G.~Tinyakov,
  JETP Lett.\  {\bf 46}, 167 (1987)
  [Pisma Zh.\ Eksp.\ Teor.\ Fiz.\  {\bf 46}, 134 (1987)].
  
\bibitem{Hawking87}
S. W. Hawking, 
Phys. Rev. D {\bf 37} (1988) 904 


\bibitem{ArkaniHamed:2007js}
  N.~Arkani-Hamed, J.~Orgera and J.~Polchinski,
  JHEP {\bf 0712} (2007) 018
  doi:10.1088/1126-6708/2007/12/018
  [arXiv:0705.2768 [hep-th]].
  
\bibitem{Bergshoeff:2005zf}
  E.~Bergshoeff, A.~Collinucci, A.~Ploegh, S.~Vandoren and T.~Van Riet,
  JHEP {\bf 0601} (2006) 061
  doi:10.1088/1126-6708/2006/01/061
  [hep-th/0510048].


\bibitem{Bergman:2007ss}
  A.~Bergman and J.~Distler,
  arXiv:0707.3168 [hep-th].


\bibitem{Coleman:1988cy}
  S.~R.~Coleman,
  Nucl.\ Phys.\ B {\bf 307} (1988) 867.
  doi:10.1016/0550-3213(88)90110-1

\bibitem{Giddings88}
  S.~B.~Giddings and A.~Strominger,
  Nucl.\ Phys.\ B {\bf 307} (1988) 854.

\bibitem{ArkaniHamed:2006dz}
  N.~Arkani-Hamed, L.~Motl, A.~Nicolis and C.~Vafa,
  JHEP {\bf 0706} (2007) 060
  doi:10.1088/1126-6708/2007/06/060
  [hep-th/0601001].

\bibitem{Montero:2015ofa}
  M.~Montero, A.~M.~Uranga and I.~Valenzuela,
  JHEP {\bf 1508} (2015) 032
  doi:10.1007/JHEP08(2015)032
  [arXiv:1503.03886 [hep-th]].

\bibitem{Hebecker:2016dsw}
  A.~Hebecker, P.~Mangat, S.~Theisen and L.~T.~Witkowski,
  arXiv:1607.06814 [hep-th].
  


\bibitem{Bergshoeff:2004pg}
  E.~Bergshoeff, A.~Collinucci, U.~Gran, D.~Roest and S.~Vandoren,
  Fortsch.\ Phys.\  {\bf 53} (2005) 990
  doi:10.1002/prop.200410227
  [hep-th/0412183].
  
  
  
\bibitem{Kachru:1998ys}
  S.~Kachru and E.~Silverstein,
  Phys.\ Rev.\ Lett.\  {\bf 80} (1998) 4855
  doi:10.1103/PhysRevLett.80.4855
  [hep-th/9802183].

\bibitem{Breitenlohner:1987dg}
  P.~Breitenlohner, D.~Maison and G.~W.~Gibbons,
  Commun.\ Math.\ Phys.\  {\bf 120} (1988) 295.
  doi:10.1007/BF01217967

\bibitem{Gutperle:2002km}
  M.~Gutperle and W.~Sabra,
  Nucl.\ Phys.\ B {\bf 647} (2002) 344
  doi:10.1016/S0550-3213(02)00942-2
  [hep-th/0206153].



\bibitem{Bergshoeff:2008be}
  E.~Bergshoeff, W.~Chemissany, A.~Ploegh, M.~Trigiante and T.~Van Riet,
  Nucl.\ Phys.\ B {\bf 812} (2009) 343
  doi:10.1016/j.nuclphysb.2008.10.023
  [arXiv:0806.2310 [hep-th]].

\bibitem{Bossard:2009at}
  G.~Bossard, H.~Nicolai and K.~S.~Stelle,
  JHEP {\bf 0907} (2009) 003
  doi:10.1088/1126-6708/2009/07/003
  [arXiv:0902.4438 [hep-th]].

\bibitem{Zamolodchikov:1986gt}
  A.~B.~Zamolodchikov,
  JETP Lett.\  {\bf 43} (1986) 730
   [Pisma Zh.\ Eksp.\ Teor.\ Fiz.\  {\bf 43} (1986) 565].

\bibitem{Belitsky:2000ws}
  A.~V.~Belitsky, S.~Vandoren and P.~van Nieuwenhuizen,
  Class.\ Quant.\ Grav.\  {\bf 17} (2000) 3521
  doi:10.1088/0264-9381/17/17/305
  [hep-th/0004186].

\bibitem{Balasubramanian:1998de}
  V.~Balasubramanian, P.~Kraus, A.~E.~Lawrence and S.~P.~Trivedi,
  Phys.\ Rev.\ D {\bf 59} (1999) 104021
  doi:10.1103/PhysRevD.59.104021
  [hep-th/9808017].
  
\bibitem{Corrado:2002wx}
  R.~Corrado, M.~Gunaydin, N.~P.~Warner and M.~Zagermann,
  Phys.\ Rev.\ D {\bf 65} (2002) 125024
  doi:10.1103/PhysRevD.65.125024
  [hep-th/0203057].
  
\bibitem{Louis:2015dca}
  J.~Louis, H.~Triendl and M.~Zagermann,
  JHEP {\bf 1510} (2015) 083
  doi:10.1007/JHEP10(2015)083
  [arXiv:1507.01623 [hep-th]].


\bibitem{Theis:2002er}
  U.~Theis and S.~Vandoren,
  JHEP {\bf 0209} (2002) 059
  doi:10.1088/1126-6708/2002/09/059
  [hep-th/0208145].
  
\bibitem{Andrianopoli:2012ee}
  L.~Andrianopoli, R.~D'Auria, P.~Giaccone and M.~Trigiante,
  JHEP {\bf 1212} (2012) 078
  doi:10.1007/JHEP12(2012)078
  [arXiv:1210.4047 [hep-th]].

\bibitem{Giddings:1989bq}
  S.~B.~Giddings and A.~Strominger,
  Phys.\ Lett.\ B {\bf 230} (1989) 46.
  doi:10.1016/0370-2693(89)91651-1
  
\bibitem{Maldacena:2004rf}
  J.~M.~Maldacena and L.~Maoz,
  JHEP {\bf 0402} (2004) 053
  doi:10.1088/1126-6708/2004/02/053
  [hep-th/0401024].
  
\bibitem{Rubakov:1996cn} 
  V.~A.~Rubakov and O.~Y.~Shvedov,
  Phys.\ Lett.\ B {\bf 383}, 258 (1996)
  doi:10.1016/0370-2693(96)00766-6
  [gr-qc/9604038].
  
 \bibitem{Hertog17}
 T. Hertog, B. Truijen, T. Van Riet, in progress 
  
\end{thebibliography}
\end{document}